\documentstyle[12pt]{article}
\relax
\begin{document}
\begin{titlepage}
\vfill

\hskip 4in {ISU-HET-97-3}

\hskip 4in {May 1997}
\vspace{1 in}
\begin{center}
{\large \bf $\Delta I=3/2$ and $\Delta S =2$ Hyperon Decays in
Chiral Perturbation Theory}\\

\vspace{1 in}
{\bf  Xiao-Gang~He$^{(a)}$} 
{\bf  and G.~Valencia$^{(b)}$}\\
{\it  $^{(a)}$ School of Physics,
               University of Melbourne,
               Parkville, Vic 3052, Australia}\\
{\it  $^{(b)}$ Department of Physics and Astronomy,
               Iowa State University,
               Ames IA 50011}\\
\vspace{1 in}
\end{center}
\begin{abstract}

We study the $|\Delta I| =3/2$ and $|\Delta S|=2$ amplitudes for 
hyperon decays of the 
form $B \rightarrow B^\prime \pi$ at lowest order in chiral perturbation 
theory. At this order, the $\Delta I=3/2$ amplitudes 
depend on only one constant. 
We extract the value of this constant from experiment and find a reasonable 
description of these processes within experimental errors. The same constant 
determines the $\Delta S =2$ transitions which, in the standard model, are 
too small to be observed. We find that new physics with parity odd 
$\Delta S=2$ interactions can produce observable rates in hyperon decays 
while evading the bounds from  $K^0 - \overline{K}^0$ mixing.

\end{abstract}

\end{titlepage}

\clearpage

\section{Introduction}

The currently available experimental information on hyperon non-leptonic 
decay amplitudes indicates that the $\Delta I =3/2$ transitions are largely 
suppressed with respect to their $\Delta I =1/2$ counterparts \cite{overseth}. 
The actual value of the $\Delta I =3/2$ amplitudes is rather uncertain. 
There is an opportunity to improve our knowledge of some of these 
amplitudes in E871, the experiment currently searching for $CP$ violation 
in hyperon decays at Fermilab \cite{proposal}.

On the theoretical side, the $\Delta I =1/2$ amplitudes have been calculated 
within the framework of chiral perturbation theory ($\chi$PT) to next to 
leading order \cite{bijnens}. It is known that at leading order it is 
possible to obtain a good fit to the $p$-wave amplitudes or the $s$-wave 
amplitudes but not to both in terms of the two couplings that appear in the 
chiral Lagrangian. At next to leading order there are enough couplings to 
fit all the amplitudes but there is no predictive power.

In this paper we study the $\Delta I =3/2$ transitions at leading order 
in $\chi$PT. At this order, there is a unique operator transforming as a 
$(27_L,1_R)$ under chiral rotations, and hence we are able to predict the 
amplitudes in terms of one coupling constant. We find that the currently 
available data (which has large uncertainties) 
is consistent with this description at the two standard deviation level. 
An improvement in the measurements would thus 
constitute a very useful test of $\chi$PT. 

Within the standard model 
it is possible to relate the $\Delta I =3/2$ amplitudes to $\Delta S =2$ 
amplitudes in hyperon decays. 
Using this relation we compute the rates for $\Delta S=2$ hyperon 
decays within the standard model. We show that it is possible for new 
$\Delta S =2$ interactions to induce hyperon decays at an observable level 
while evading the bounds from $K^0-\overline{K}^0$ mixing.

In $\chi$PT the pseudo-Goldstone boson fields are incorporated via the 
matrix:
\begin{eqnarray}
\phi &=& {1 \over \sqrt{2}}
\left( \begin{array}{ccc}
\pi^0 /\sqrt{2}+\eta /\sqrt{6} & \pi^+ & K^+ \\
\pi^- &-\pi^0 /\sqrt{2}+\eta /\sqrt{6} & K^0 \\
K^- & \overline{K^0} & -2 \eta /\sqrt{6}
       \end{array} \right)
\label{pions}
\end{eqnarray}
into the matrix $\Sigma = \exp(i2\phi/f_\pi)$ that transforms 
as $\Sigma \rightarrow L \Sigma R^\dagger$ under chiral 
$SU(3)_L\times SU(3)_R$. The baryon octet is introduced via 
the matrix:
\begin{eqnarray}
B &=& 
\left( \begin{array}{ccc}
\Sigma^0 /\sqrt{2}+\Lambda /\sqrt{6} & \Sigma^+ & p \\
\Sigma^- &-\Sigma^0 /\sqrt{2}+\Lambda /\sqrt{6} & n \\
\Xi^- & \Xi^0 & -2 \Lambda /\sqrt{6}
       \end{array} \right)
\label{baryons}
\end{eqnarray}
transforming as $B \rightarrow U B U^\dagger$. The matrix $U$ is, in turn, 
defined by the transformation properties of $\xi=\exp(i\phi/f_\pi)$, which 
under an $SU(3)_L\times SU(3)_R$ transformation goes into 
$\xi \rightarrow L\xi U^\dagger = U \xi R^\dagger$. In terms of these 
fields one easily constructs the 
lowest order chiral Lagrangian that describes the strong interactions of 
the baryon and pseudoscalar meson octets \cite{georgi,dogoho}. It is given by:
\begin{eqnarray}
{\cal L}_S &=& {f_\pi^2 \over 4}{\rm Tr}\bigl( \partial_\mu \Sigma 
\partial^\mu \Sigma^\dagger \bigr) + i{\rm Tr}\bigl(\overline{B}\not{\partial}
B\bigr)-M{\rm Tr}\overline{B}B +{i\over 2}{\rm Tr}\overline{B}\gamma_\mu
\bigl[\xi\partial^\mu \xi^\dagger+\xi^\dagger\partial^\mu \xi,B\bigr]
\nonumber \\
&+&i{D\over 2}{\rm Tr}\overline{B}\gamma_\mu \gamma_5
\bigl\{\xi\partial^\mu \xi^\dagger-\xi^\dagger\partial^\mu \xi,B\bigr\}
+i{F\over 2}{\rm Tr}\overline{B}\gamma_\mu \gamma_5
\bigl[\xi\partial^\mu \xi^\dagger-\xi^\dagger\partial^\mu \xi,B\bigr]
\label{strongl}
\end{eqnarray}
In our normalization $f_\pi \approx 93$~MeV, and we use the values 
of $D=0.80\pm 0.14$ and 
$F=0.50\pm 0.12$ obtained in a lowest order fit to hyperon semileptonic 
decays in Ref.~\cite{bijnens,manjen}. To go beyond leading order in $\chi$PT 
it is more convenient to use a formulation in which the baryons 
are treated as heavy static fields \cite{manjen,springer}. In this paper we do 
not attempt to go beyond leading order and will use Eq.~\ref{strongl}.

\section{$|\Delta I| = 3/2$ Amplitudes}

Within the standard model the $\Delta I =3/2$ transitions are induced 
by an effective Hamiltonian that transforms as a $(27_L,1_R)$ under chiral 
rotations: 
\begin{equation}
{\cal H}_{eff}^{(27_L,1_R)} = 
{G_F \over \sqrt{2}}V^*_{ud}V_{us}\bigl({c_1+c_2 \over 3}\bigr)
{\cal O}^{(27_L,1_R)}_{\Delta I =3/2} \, +\, h.c.
\label{weaktsl}
\end{equation}
The four-quark operator ${\cal O}(\Delta I =3/2)$ is 
$4~ T^{jk}_{lm}\overline{\psi}^j_L\gamma_\mu\psi^l_L
\overline{\psi}^k_L\gamma^\mu\psi^m_L$, 
and at lowest order in $\chi$PT (${\cal O}(p^0)$) it has a unique 
chiral realization in terms of a coupling that we call $b_{27}$ given by:
\begin{equation}
{\cal O}^{(27_L,1_R)}_{\Delta I = 3/2} = 
m_\pi^2f_\pi b_{27} T^{jk}_{lm}
\bigl(\xi\overline{B}\xi^\dagger\bigr)^l_j
\bigl(\xi B\xi^\dagger\bigr)^m_k
\label{dichiral}
\end{equation}
The short distance Wilson coefficients $c_1 = -0.58$, $c_2 =1.31$ have been 
calculated in Ref.~\cite{buras} (we follow their notation and take 
$\mu =1~GeV$, $\Lambda_{QCD} =200~MeV$). 
The nonvanishing components of the tensor $T^{jk}_{lm}$ are 
$T^{1,2}_{1,3}=T^{2,1}_{1,3}=T^{1,2}_{3,1}=T^{2,1}_{3,1}=1/2$, 
and $T^{2,2}_{2,3}=T^{2,2}_{3,2}=-1/2$ \cite{georgi}.

It is conventional to write the invariant matrix element for the decays 
$B_i \rightarrow B_f \pi$ in the form:
\begin{equation}
{\cal M} = G_F m_\pi^2 \overline{U}_f\bigl(A+B\gamma_5\bigr)U_i
\label{matel}
\end{equation}
The $A$ amplitudes follow directly from Eq.~\ref{dichiral}, whereas 
the $B$ amplitudes are obtained from the baryon pole diagrams in 
which Eq.~\ref{dichiral} is responsible for a baryon to baryon transition 
combined with pion emission from one of the baryon lines 
according to Eq.~\ref{strongl}. We present our results in 
Table~\ref{t; di}.

\begin{table}[htb]
\centering
\caption{Matrix elements of ${\cal O}^{(27_L,1_R)}_{\Delta I = 3/2}$ for 
nonleptonic hyperon decays in units $m^2_\pi b_{27}$.}
\begin{tabular}{|c|c|c|} \hline
Decay mode & $A$ & $B$ \\ \hline & & \\
$\Lambda \rightarrow p \pi^-$ & 0 & ${1\over 2\sqrt{3}}D
{M_\Lambda+M_N \over M_\Sigma - M_N}$ \\
$\Lambda \rightarrow n \pi^0$ & 0 & ${1\over \sqrt{6}}D
{M_\Lambda+M_N \over M_\Sigma - M_N}$ \\
$\Sigma^+ \rightarrow n \pi^+$ & $-{3\over 2\sqrt{2}}$ & $-{1\over 2\sqrt{2}}
(3F+D) {M_\Sigma+M_N \over M_\Sigma - M_N}$ \\
$\Sigma^+ \rightarrow p \pi^0$ & ${1\over 4}$ & ${1\over 4}
(F-D) {M_\Sigma+M_N \over M_\Sigma - M_N}$ \\
$\Sigma^- \rightarrow n \pi^-$ & ${1\over \sqrt{2}}$ & ${1\over \sqrt{2}}
F {M_\Sigma+M_N \over M_\Sigma - M_N}$ \\
$\Xi^0 \rightarrow \Lambda \pi^0$ & 0 & $-{1\over \sqrt{6}}D
{M_\Xi+M_\Lambda \over M_\Xi- M_\Lambda}$ \\
$\Xi^- \rightarrow \Lambda \pi^-$ & 0 & $-{1\over 2\sqrt{3}}D
{M_\Xi+M_\Lambda \over M_\Xi- M_\Lambda}$ \\
& & \\ \hline
\end{tabular}
\label{t; di}
\end{table}

We construct ratios of $\Delta I =3/2$ amplitudes as calculated with the 
aid of Table~\ref{t; di} to $\Delta I =1/2$ amplitudes taken from 
experiment, and compare with the value of these ratios as extracted 
from experiment in Ref.~\cite{overseth}. In the notation of 
Ref.~\cite{overseth}, we present this comparison in Table~\ref{t; comp}.

\begin{table}[htb]
\centering
\caption{Comparison of lowest order $\chi$PT predictions and experiment 
for $(\Delta I =3/2)/(\Delta I =1/2)$ amplitude ratios.}
\begin{tabular}{|c|c|c|c|} \hline
Process & Theory from Eqs.~\ref{weaktsl}, \ref{dichiral} & 
Experiment from Ref.~\cite{overseth} & $b_{27}$  \\ \hline & & & \\
$\Lambda$ decay & & & \\
$A_3/A_1$ & $0$ & $0.027 \pm 0.008$ & $-$ \\
$B_3/B_1$ & $-0.010 b_{27}$ & $0.030 \pm 0.037$ & $-3.0 \pm 3.7$ \\
\hline & & & \\ $\Xi$ decay & & & \\
$A_3/A_1$ & $0$ & $-0.046 \pm 0.014$ & $-$ \\
$B_3/B_1$ & $-0.029 b_{27}$ & $-0.01 \pm 0.04$ & $0.3 \pm 1.4$ \\
\hline & & &\\  $\Sigma$ decay & & & \\
$A_3/A_-$ & $0.023 b_{27}$ & $-0.061 \pm 0.024$ & $-2.7 \pm 1.1$ \\
$B_3/B_+$ & $0.009 b_{27}$ & $-0.074 \pm 0.027$ & $-7.9 \pm 2.9$ \\
& & & \\ \hline
\end{tabular}
\label{t; comp}
\end{table}

From the comparison in Table~\ref{t; comp} we see that there is 
qualitative agreement between the data and the predictions. For 
example, the ratios for $A$ amplitudes in $\Lambda$ and $\Xi$ decays 
do not occur at lowest order in $\chi$PT and are thus expected to 
be about three times smaller than the corresponding ratio in 
$\Sigma$ decays (the typical size of next order corrections $m_s/M_N$). 
In the last column of Table~\ref{t; comp} we present the fit to 
$b_{27}$ from the four ratios that do not vanish at leading order 
in $\chi$PT. We see that these values are consistent at the two 
standard deviation level. It is clear, however, that a truly 
quantitative comparison will not be possible until the experimental 
uncertainties are significantly reduced.

The $\Delta I =1/2$ amplitudes calculated in $\chi$PT are known to have 
large corrections at next to leading order \cite{bijnens}. We can get 
a simple estimate for the size of these corrections in the $\Delta I =3/2$ 
sector by examining one of the possible next to leading order (${\cal O}(p)$) 
chiral realizations of ${\cal O}^{(27_L,1_R)}_{\Delta I =3/2}$. We choose one 
that reproduces the factorized current-current 
form that appears in vacuum saturation where all the coefficients are known: 
\begin{equation}
{\cal O}^{(27_L,1_R)}_{\Delta I =3/2} = 
m_\pi^2 f_\pi{\tilde{b}_{27} \over M_N} T^{i,j}_{k,l}
\bigl[
\bigl(\xi\overline{B}\bigr)^a_i \gamma_\mu \bigl(B\xi^\dagger\bigr)^k_a -
\bigl(\overline{B}\xi^\dagger\bigr)^k_a \gamma_\mu \bigl(\xi B \bigr)^a_i \bigr]
\bigl(\Sigma \partial^\mu \Sigma^\dagger)^l_j
\label{vacsat}
\end{equation}
This operator gives contributions to the $A$ amplitudes that would be 
suppressed 
with respect to the leading order ones from Eq.~\ref{dichiral} by factors 
of $m_s/M_N \sim 30\%$ if $\tilde{b}_{27} \sim b_{27}$ as suggested by 
naive power counting. Instead we find in vacuum saturation that 
$\tilde{b}_{27} = -8/3f_\pi M_N/m^2_\pi \approx -12$, two to six times 
larger than our fit for $b_{27}$. It seems likely that next to leading 
corrections in the $\Delta I =3/2$ sector are as important as in the 
$\Delta I =1/2$ sector. With improved experimental accuracy, 
a next to leading order analysis will become necessary to test the 
predictions of $\chi$PT.

\section{$|\Delta S| = 2$ Amplitudes}

It is possible to relate the coupling responsible for the $\Delta I
=3/2$ transitions, $b_{27}$, to the coupling responsible for the $\Delta S =2$ 
transitions within the standard model. This is entirely analogous to the 
relation between the $B$-parameter in $K^0 -\overline{K}^0$ mixing and 
the $\Delta I =3/2$ $K^+ \rightarrow \pi^+ \pi^0$ decay \cite{donoghue}. 

The effective $\Delta S =2$ interactions within the standard model are 
given by:
\begin{eqnarray}
{\cal H}^{\Delta S =2}_{box} &=& {G_F^2 \over 16 \pi^2}\bigl[
\bigl(V^*_{cd}V_{cs}\bigr)^2H(x_c)m^2_c\eta_1 +
\bigl(V^*_{td}V_{ts}\bigr)^2H(x_t)m_t^2\eta_2 \nonumber \\ &+&
2 \bigl(V^*_{td}V_{ts}V^*_{cd}V_{cs}\bigr)\overline{G}(x_c,x_t)m_c^2\eta_3\bigr]
{\cal O}^{27}_{\Delta S =2} \,+\, h.c. 
\label{dssham}
\end{eqnarray}
The QCD correction factors $\eta_i$ have been calculated in 
Ref.~\cite{buras}. The lowest order chiral realization of the 
$\Delta S=2$ operator is given by:
\begin{equation}
{\cal O}^{(27_L,1_R)}_{\Delta S =2}= 
m_\pi^2 f_\pi b_{27} T^{jk}_{lm}
\bigl(\xi\overline{B}\xi^\dagger\bigr)^l_j
\bigl(\xi B\xi^\dagger\bigr)^m_k 
\label{dsschiral}
\end{equation}
where, in this case, the relevant component 
of $T^{jk}_{lm}$ is $T^{2,2}_{3,3}=1$. 

Since the coupling $b_{27}$ has been extracted in the previous section 
we can predict the $\Delta S =2$ {\it short distance} amplitudes within 
the standard model. We present the matrix elements of 
${\cal O}^{(27_L,1_R)}_{\Delta S =2}$ in Table~\ref{t; dss}. 
\begin{table}[htb]
\centering
\caption{Matrix elements of ${\cal O}^{(27_L,1_R)}_{\Delta S =2}$ for 
$\Delta S =2$ nonleptonic hyperon decays in units of $m^2_\pi b_{27}$.} 
\begin{tabular}{|c|c|c|} \hline
Decay mode & $A$ & $B$ \\ \hline &  &\\ 
$\Xi^0 \rightarrow p \pi^-$ & $-{1\over \sqrt{2}}$ & $-{1\over \sqrt{2}}(D+F)
{M_\Xi+M_N \over M_\Xi- M_N}$ \\
$\Xi^0 \rightarrow n \pi^0$ & $1$ & $F
{M_\Xi+M_N \over M_\Xi- M_N}$ \\
$\Xi^- \rightarrow n \pi^-$ & $-{1\over \sqrt{2}}$ & ${1\over \sqrt{2}}(D-F)
{M_\Xi+M_N \over M_\Xi- M_N}$ \\
& & \\ \hline
\end{tabular}
\label{t; dss}
\end{table}

These amplitudes lead to branching ratios of the order of $10^{-17}$, far 
too small to be measured in the foreseeable future. 
We should point out that there are also long distance 
contributions to these amplitudes obtained with the $\Delta S=1$ weak 
Hamiltonian acting twice; they are also unobservably small.

It is interesting to ask whether it is possible to observe a $\Delta S =2$ 
non-leptonic hyperon decay in light of the constraints that arise 
from $K^0- \overline{K}^0$ mixing. 
This will only be possible if the new interaction has a contribution to 
$K^0- \overline{K}^0$ mixing that is 
suppressed with respect to its contribution 
to hyperon decays by at least a factor of $G_F m^2_K$. This is possible if 
the low energy effective Hamiltonian for the new $\Delta S =2$ interactions  
contains only parity odd operators with couplings of order $G_F$ or less. In 
this case the new operators will not contribute 
directly to $K^0- \overline{K}^0$ mixing. They may induce a long distance 
contribution (at one-loop),  
when combined with a parity-odd vertex from the $\Delta S =0$ weak interaction. 
This contribution, however, 
will be at most of the same order as the standard model short distance 
contribution and the new operator is not significantly constrained. 
On the other hand, this type of operator can contribute at the weak level to 
the s-wave $A$ amplitudes in hyperon decays. One such operator at the 
quark level is 
\begin{equation}
{\cal L}_{new} = G_F {\alpha_{new} \over \alpha} \overline{d}\gamma_\mu s
\overline{d}\gamma_\mu \gamma_5  s \,+\, h.c.
\label{newdss}
\end{equation}
The factor $\alpha_{new}/\alpha$ parameterizes the strength of the new 
interactions at high energy relative to the electro-weak interactions. 
As discussed above, if we take $\alpha_{new}/\alpha \leq 1$, there will be 
no significant constraints from $K^0- \overline{K}^0$ mixing. 

Under $SU(3)_L\times SU(3)_R$ the new operator transforms as a 
$(27_L,1_R)-(1_L,27_R)$ and therefore we can easily construct its 
chiral realization in terms of the same tensor of Eq.~\ref{dsschiral}:
\begin{equation}
{\cal L}_{new} = {G_F \over 4} {\alpha_{new}\over \alpha}m^2_\pi f_\pi b_{27} 
T^{jk}_{lm}\biggl(
\bigl(\xi\overline{B}\xi^\dagger\bigr)^l_j
\bigl(\xi B\xi^\dagger\bigr)^m_k -
\bigl(\xi^\dagger\overline{B}\xi\bigr)^l_j
\bigl(\xi^\dagger B\xi\bigr)^m_k \biggr) \,+\, h.c. 
\label{dssnewchiral}
\end{equation}
This operator contributes only to the $A$ amplitudes of the decays listed 
in Table~\ref{t; dss} due to its parity odd nature. 
Its matrix elements are just twice those listed in 
Table~\ref{t; dss} for the $(27_L,1_R)$ operator\footnote{ 
In general the coefficient of this chiral Lagrangian is not 
equal to $b_{27}$ due to the mixing 
of the new high energy operator with the standard model operators. 
However, we expect them to be of the same order of magnitude so we write 
$b_{27}$ in Eq.~\ref{dssnewchiral} for clarity and assume that any difference 
is absorbed into the definition of $\alpha_{new}$.}. 
The branching ratios for these $\Delta S =2$ 
decays are then:
\begin{eqnarray}
B(\Xi^0 \rightarrow p \pi^-) & = & 0.9
\biggl({\alpha_{new}\over \alpha}\biggr)^2    \nonumber \\
B(\Xi^0 \rightarrow n \pi^0) & = &    1.8 
\biggl({\alpha_{new}\over \alpha}\biggr)^2   \nonumber \\
B(\Xi^- \rightarrow n \pi^-) & = &    1.6
\biggl({\alpha_{new}\over \alpha}\biggr)^2  
\label{ratesdss}
\end{eqnarray}
The current upper limits $B(\Xi^0 \rightarrow p \pi^-) < 1.9 \times 10^{-5}$ 
and $B(\Xi^- \rightarrow n \pi^-) < 4 \times 10^{-5}$ \cite{pdb}, imply a bound 
$\alpha_{new}/\alpha < 5 \times 10^{-3}$ (we use $b_{27}= -3$). 
It should be possible for E871 to improve this bound.

\section{Conclusions}

We have computed the $\Delta I =3/2$ amplitudes for weak non-leptonic 
hyperon decays at leading order in $\chi$PT in terms of one coupling, 
$b_{27}$. We find 
our predictions to be in qualitative agreement with experiment, but 
it will be necessary to significantly improve the 
measurements for a more meaningful quantitative comparison. 

In terms of the constant $b_{27}$ we predict the rates for $\Delta S=2$ 
non-leptonic hyperon decays and find them to be unobservably small within 
the standard model. We show that a new interaction that induces only 
parity odd $\Delta S=2$ operators is not constrained by $K^0-\overline{K}^0$ 
mixing and may be tested in hyperon decays.

\vspace{1in}

\noindent {\bf Acknowledgements} The work of G.V. was supported in
part by the DOE OJI program under contract number DEFG0292ER40730. 
The work of X-G.H. was supported by the Australian Research 
Council, and he thanks the high energy group at ISU for their 
hospitality while part of this work was performed. 
We are grateful to John F. Donoghue for useful conversations.

\end{document}